# Bifurcations and Spatial Chaos in an Open Flow Model

Frederick H. Willeboordse* and Kunihiko Kaneko†
*Department of Pure and Applied Sciences, University of Tokyo, Komaba, Tokyo 153, Japan*

It is shown that a coupled map model for open flow may exhibit spatial chaos and spatial quasiperiodicity with temporal periodicity. The locations of these patterns, which cover a substantial part of parameter space, are indicated in a comprehensive phase diagram. In order to analyze the encountered phenomena, a novel class of spatial maps is introduced which is very efficient in accurately reproducing the original spatial patterns. It is found that temporally period one spatial chaos is convectively unstable, and that it is possible to predict an essential aspect of the bifurcation behavior of the coupled system solely by considering its corresponding spatial map.

For the understanding of many complex systems in nature, the study of simple chaotic paradigms has proven to be an invaluable tool. Sometimes, substantial insight can be gained by only considering temporal aspects. In many cases, however, it is imperative to also explicitly take space into consideration. In recent years this has led to great interest in the dynamics of coupled map lattices (CML) which are not only spatio-temporal, but can also be classified by a distinct number of universality classes. Particularly, coupled logistic maps have been studied widely for their combination of computational efficiency and phenomenological richness [1-4].

In this letter, we will investigate the nature of spatial chaos by introducing a class of spatial maps which can accurately reproduce the spatial patterns of the one-way coupled logistic lattice (OCLL) in parameter regions where the latter is temporally periodic (for other work related to spatial chaos see e.g., [5] or [6]). The OCLL, due to its conceptual reminiscence with open fluid flow often referred to as a model for open flow, is defined as:

$$x^i_{n+1} = (1-\epsilon)f(x^i_n) + \epsilon f(x^{i-1}_n), \qquad (1)$$

where the local element is the logistic map $f(x_n) = 1 - \alpha x_n^2$. The parameters are the nonlinearity $\alpha$ and the coupling constant $\epsilon$, while $i$ is the index for the lattice sites, and $n$ the discrete time.

Many interesting features of (1), like spatial period-doubling and selective amplification of noise, were already reported in [7] and [8–10], but those investigations did not take the full range of possible coupling constants ($0.0 \le \epsilon \le 1.0$) into account and were generally limited to $\epsilon \lesssim 0.5$. Recent studies of the diffusively coupled logistic lattice, of which the OCLL is in principle nothing but a version with a maximally asymmetric coupling, however, have shown that larger values of $\epsilon$ may yield unexpected phenomena like the traveling wave or suppression of supertransient chaos at high nonlinearity [11], [12].

In the OCLL, for larger values of $\epsilon$, we have found the fascinating novelty of spatial chaos with temporal periodicity (see also [13]). Consequently, with regard to its spatio-temporal properties, the dynamics of the OCLL for nonlinearities past the accumulation point of the single logistic map can be described as dominantly temporal ($0.0 \le \epsilon \lesssim 0.05$), spatio-temporal ($0.05 \lesssim \epsilon \lesssim 0.45$) or spatial ($0.45 \lesssim \epsilon \le 1.0$).

A rough phase diagram for the OCLL is given in fig. 1, where the basic types of patterns we would like to distinguish here are marked as: spatially chaotic and temporally periodic patterns (SC), spatially (and temporally) periodic patterns (SP), spatially quasiperiodic patterns with temporal periodicity (SQP), and spatio-temporal patterns (STP). The STP region includes various kinds of patterns like pattern selection with remnant chaos, spatio-temporal intermittency and a large area of spatio-temporal chaos (STC). The region marked as 'ZZ' is the well-known zigzag pattern which is spatially and temporally period 2, and therefore a special case of SP. Throughout, the left boundary was fixed to 0 (the exact value does not seem to matter much).

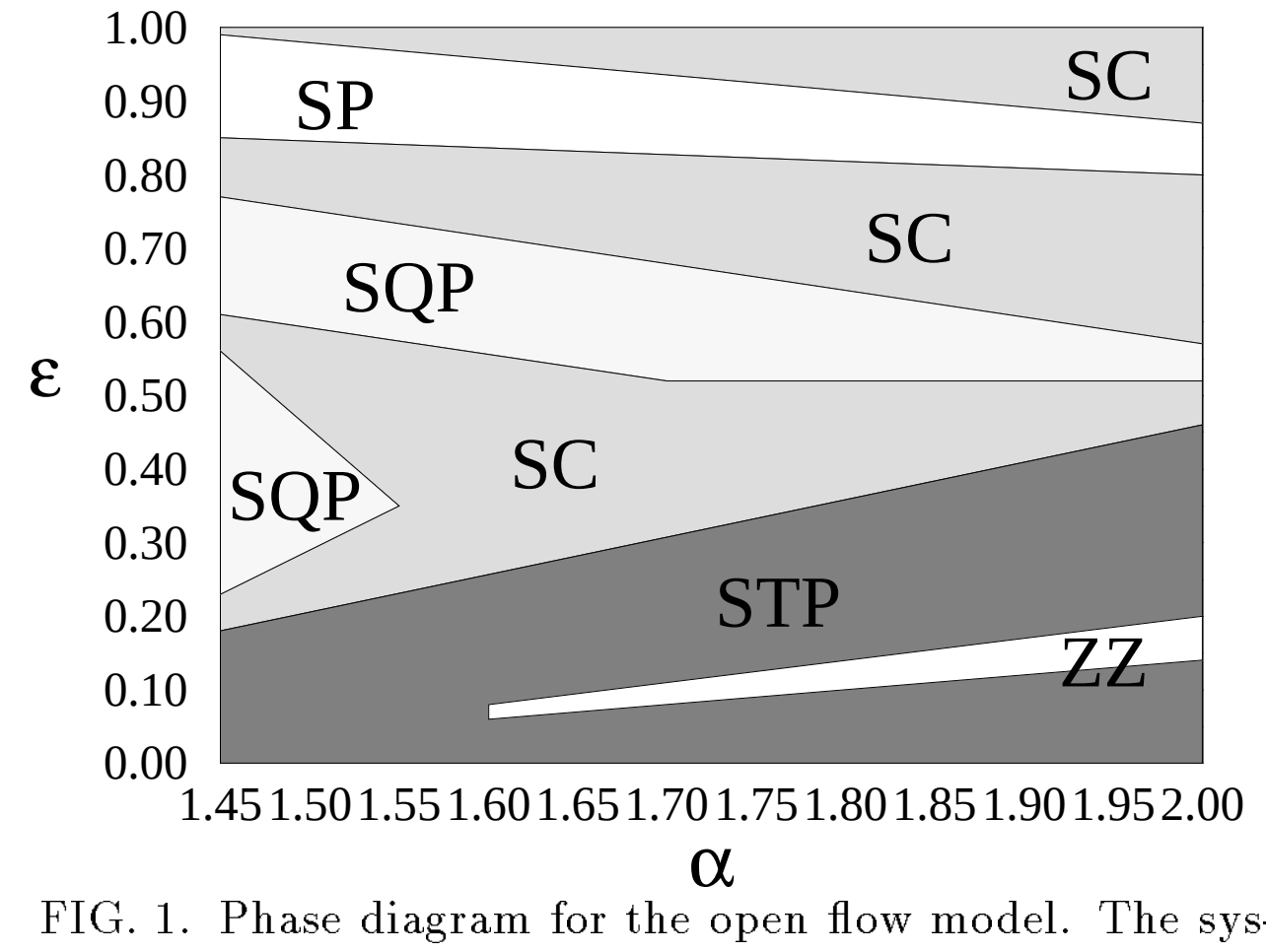

FIG. 1. Phase diagram for the open flow model. The system size is $N = 383$.

In order to illustrate the two temporally periodic and spatially non-periodic patterns, space-amplitude plots of SC and SQP are displayed in figs. 2 and 3 respectively. In the case of SC (of which a definition will be given below in terms of the newly introduced spatial map), the temporal periodicity rapidly bifurcates to 32 but then remains constant until the end of the lattice. In the case of SQP, after the initial bifurcations, the pattern has a temporal periodicity of 8 with a spatial period close to 14, while the points of the return map lie on an eye-shaped loop indicating quasiperiodicity.

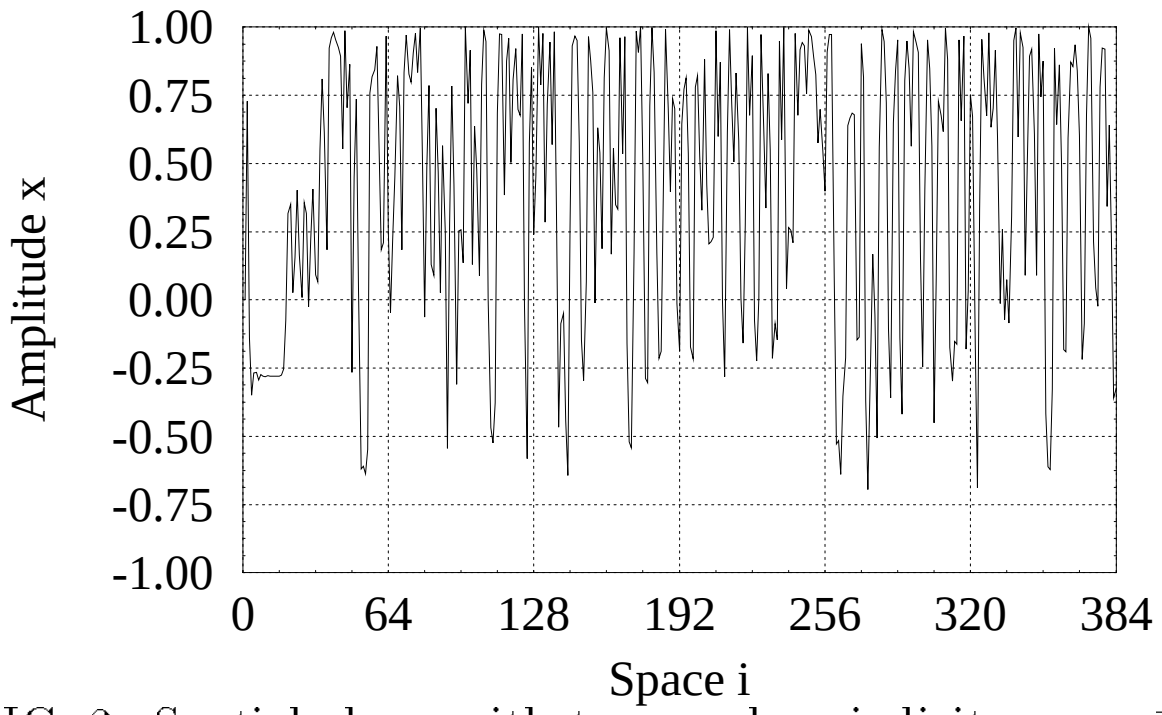

FIG. 2. Spatial chaos with temporal periodicity. $\alpha = 1.7$, $\epsilon = 0.45$ and $N = 384$. The temporal periodicities are 1 for $i = 1 - 2$, 2 for $i = 3 - 16$, 4 for $i = 17 - 31$, 8 for $i = 32 - 78$, 16 for $i = 79 - 127$ and 32 for $i = 128 - 384$ respectively.

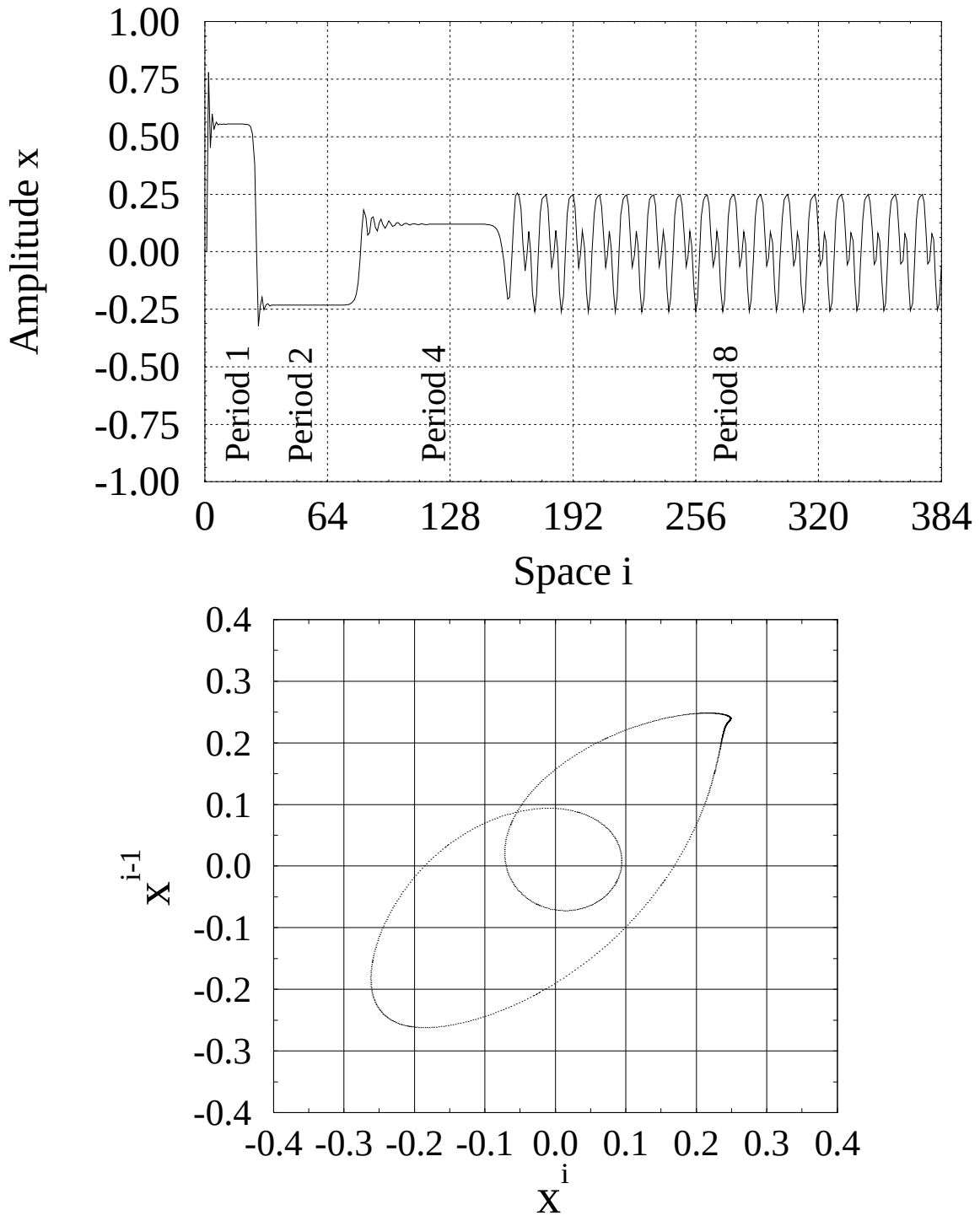

FIG. 3. Spatial quasiperiodicity with temporal periodicity. a) $\alpha = 1.45$, $\epsilon = 0.5$ and $N = 1024$. Only the first 384 sites are shown, the remaining sites are the same. The indicated periodicities are temporal periodicities. b) Return map corresponding to (a) for sites $i = 256 - 1024$.

The various areas in the phase diagram are separated by sharp lines in order to give some impression of their location. In fact, however, the lines are not very sharp at all and in many cases multiple attractors belonging to different basic patterns exist for the same parameters. In some cases this might be related to the dependence of the spatial locations of the temporal bifurcation points on the initial conditions, and thus mainly reflect the difference between up- and downflow. In other cases, however, different types of patterns may exist for the same temporal periodicity, indicating genuine multiple states. The diagram furthermore only indicates the most common patterns in a larger area of parameter space. That is to say, small SQP regions within the SC region,e.g., were not taken into consideration.

In order to analyze the phenomenon of purely spatial chaos, we employ the one-way coupling and the temporal periodicity to introduce a spatial map as an implicit equation of equal-time spatial variables. If we define $F(x_n^i) = (1-\epsilon)f(x_n^i) + \epsilon f(x_n^{i-1})$, then a lattice site $x_n^i$ which has a temporal periodicity $k$ must fulfill,

$$x_n^i = F^k(x_n^i), \tag{2}$$

where $F^k(x_n^i)$ is the $k$-the iterate of $F(x_n^i)$. We can now formally define a spatial map corresponding to eq. (2) in an implicit form as

$$G^k(x^i) = -x^i + F^k(x^i), \tag{3}$$

with which a lattice can be generated by successively finding the roots of eq. (3) and incrementing the index $i$. Since eq. (3) depends on $k$ variables, $x^{i-1}, \dots, x^{i-k}$ need to be supplied as initial conditions (note the absence of the time-index $n$). Thus we can reduce even an infinite lattice, and hence an infinite dimensional system, to a $k$-dimensional map, while substantially improving the opportunities for mathematical and numerical analysis. In general, of course, an equation like (2) will have a large number of solutions, especially if the periodicity $k$ is high. We will now show, however, that for sufficiently large $\epsilon$ there is only one.

The derivative of eq. (3) is given by:

$$G^{k\prime}(x^i) = -1 + (1-\epsilon)^k(-2\alpha)^k \prod_{m=0}^{m=k-1} K^m(x^i), \tag{4}$$

where $K^m(x^i)$ is the $m$-th iterate of $K(x^i) = (1-\epsilon)f(x^i) + \epsilon f(x^{i-1})$ and $K^0(x^i) = x^i$. If $G^{k\prime}(x^i) < 0 \forall x^i$, there is at most one root. Since $|\prod_{m=0}^{m=k-1} K^m(x^i)| \le 1$ this will be the case if $(1-\epsilon)^k(-2\alpha)^k < 1$, i.e., if

$$\epsilon > 1 - \frac{1}{2\alpha}. \tag{5}$$

Thus, the spatial map (3) exactly represents the original model (1) as long as condition (5) holds (it should be noted that this condition covers a much larger area than the region which has a temporal periodicity of one in the open flow model, and that it also holds for $k \to \infty$ in (3)).

Numerically, it was furthermore verified that for basically the entire temporally periodic region of the OCLL, the corresponding spatial maps had only one root [13], while the STP area yielded a great multiplicity of roots. We therefore believe that the arise of spatio-temporal patterns below $\epsilon_{min}(\alpha)$ may be associated with the occurrence of multiple roots in the spatial map which prevent the system from reaching a periodic state.

With the help of our spatial map, spatial chaos can formally be defined by the the requirement that $\lambda_G > 0$, where $\lambda_G$ is the spatial lyapunov exponent of eq. (3) (the temporal Lyapunov exponent for a temporally periodic lattice site will of course be negative, see also below).

In most cases, an analytical solution for eq. (3) cannot be found. For $k = 1$, however, it is a quadratic equation, and the root is given by (the other root is outside of the allowed domain)

$$x^i = \frac{-1 + \sqrt{1 + 4(1-\epsilon)\alpha(1 - \epsilon\alpha(x^{i-1})^2)}}{2(1-\epsilon)\alpha}. \tag{6}$$

This forms a spatial map, which, contrary to eq. (3), is of an explicit nature and therefore extremely efficient in generating the associated spatial patterns.

For values of the coupling constant $\epsilon$ where the OCLL in principle has a temporal periodicity of one, and where thus the spatial map (6) is a valid representation of the OCLL, eq. (6) undergoes a bifurcation cascade to chaos as is shown in fig. 4.

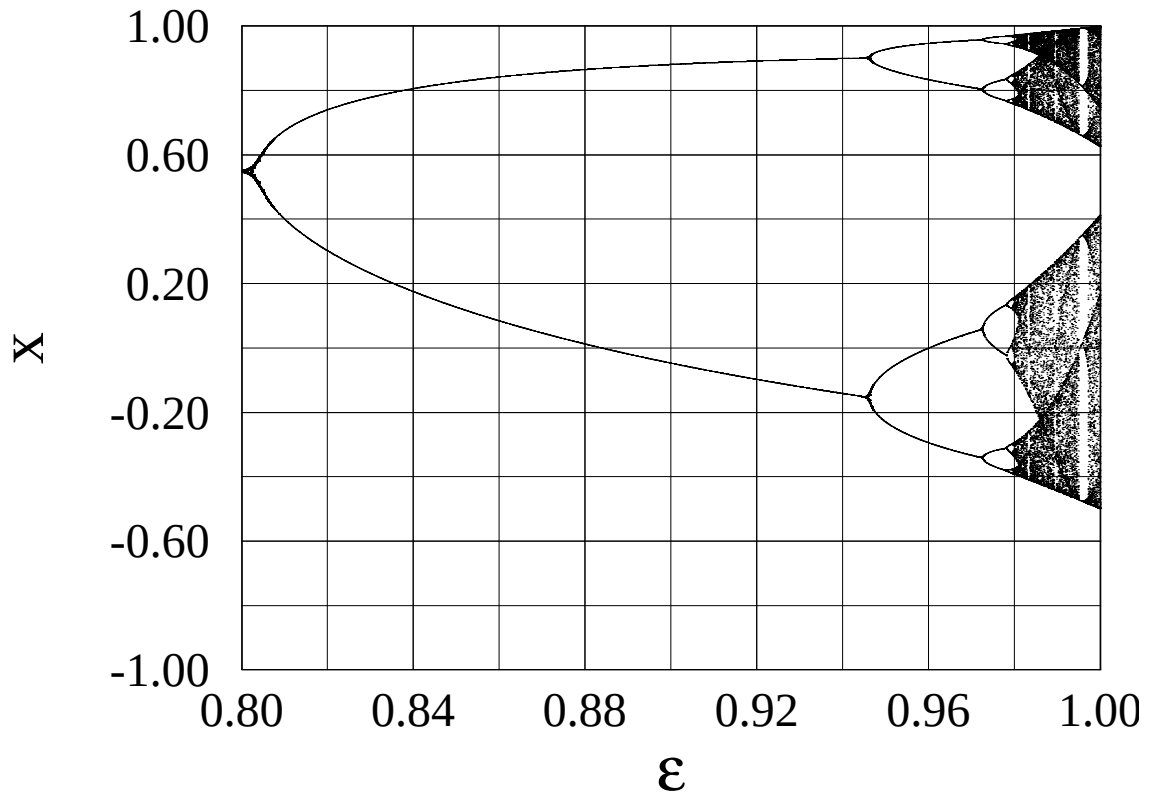

FIG. 4. Bifurcation cascade of the $k = 1$ spatial map (6). The nonlinearity is $\alpha = 1.5$.

These bifurcations are strictly in the spatial direction, and as such do not affect the temporal periodicity of the original model (1), and accordingly the possibility of spatial chaos is a natural consequence. It should be noted here that the spatial bifurcations mentioned above are fundamentally different from the spatial period doubling reported in [7]. In the former case, the temporal periodicity does not change, while a parameter does, and in the latter case, the period doubling occurs in the spatial direction and coincides with a doubling of the temporal periodicity, while the parameters remain the same.

With the help of our spatial map, we can accurately reproduce the temporally periodic spatial patterns of the OCLL. In the OCLL, however, the temporal periodicity is not necessarily a constant throughout the lattice. If we consider fig. 3, e.g., there are three homogeneous sections with temporal periodicities of 1, 2 and 4 respectively, and the spatially quasiperiodic pattern has a temporal periodicity of 8. Indeed, the $k = 1$ spatial map yields exactly the first section, the $k = 2$ the second, etc..

From this two important questions arise: why does the OCLL bifurcate temporally, despite the existence of a stable and accurate solution in the spatial map, and, is it possible to use the data from the spatial map to predict whether the OCLL will undergo temporal bifurcations or not?

These questions are naturally related to the one of stability, and we will now show that both can be answered by considering co-moving Lyapunov exponents [8] with regard to the spatial map. Conventionally, the spectrum of Lyapunov exponents would be determined by taking the logarithm of the eigenvalues of the product of Jacobi matrices, which, in this case would yield:

$$\lambda^i = \log(1-\epsilon) + \frac{1}{T}\sum_{n=1}^{n=T} \log f'(x_n^i), \tag{7}$$

where $T \to \infty$. For large $\epsilon$, the $\log(1-\epsilon)$ will cause all the Lyapunov exponents to be negative and there seems to be no reason for the lattice to temporally bifurcate at some point downflow. A different approach is therefore necessary and the distinction between absolute and convective stability which were shown to be essential concepts for the understanding of open flow [8] and which have no direct counterpart in the spatial maps, needs to be taken into account. A system is called absolutely unstable if it is unstable in any frame, and convectively unstable if it is only unstable in some moving frame.

Let us consider the $k = 1$ case. In a similar way as for the homogeneous fixed point state [2], it can be argued that, for large $\epsilon$, the Lyapunov exponent in the co-moving frame that yields the maximum growth of an initial perturbation in the OCLL may be obtained as [13]

$$L_{\max} = \frac{1}{n}\sum_{i=0}^{i=n-1} \log|f'(x^i)|. \tag{8}$$

We can then generate a spatial pattern with the spatial map and use eq. (8) to determine the pattern's stability in the co-moving frame. Our numerical simulations indicate that if $L_{\max} < 0$, the pattern is absolutely stable, and a valid solution of the OCLL for all system sizes. If, however, $L_{\max} > 0$ (while $\lambda^i < 0$) the pattern generated by the spatial map is convectively unstable, and a valid solution of the OCLL only for those lattice sites which have the same temporal periodicity.

In fig. 5, the Lyapunov exponent according to eq. (8) is plotted versus the coupling constant $\epsilon$ together with the regular exponent for eq. (6). In both cases, the necessary values of $x^i$ were obtained with the help of the spatial map (6). Positive co-moving Lyapunov exponents can clearly be seen for parameter values where the $k = 1$ spatial map is periodic, hence implying regions of convective instability. Our numerical results indicate that this corresponds to states of the OCLL which will temporally bifurcate. In this way, the spatial map provides us with an excellent tool to predict whether the original one-way coupled logistic lattice will bifurcate or not.

Since the Lyapunov exponent in the co-moving frame is larger than the one in the stationary frame, it is implied that for the temporal period one case, true spatial chaos does not exist. However, the dips between the bifurcation points, corresponding to superstable orbits which also exist in the periodic windows, extend to minus infinity, guaranteeing ample stability in their neighborhoods to conclude that spatial patterns arbitrarily close to chaos can stably exist in the one-way coupled logistic lattice. Thus far, our numerical results also indicate that the correspondence between temporal bifurcations in the OCLL and $L_{\max} > 0$ holds for $k > 1$. Indeed, regardless of the temporal periodicity, all the spatially chaotic patterns seem to yield $L_{\max} > 0$ indicating that the instability of true spatial chaos might very well be a general feature.

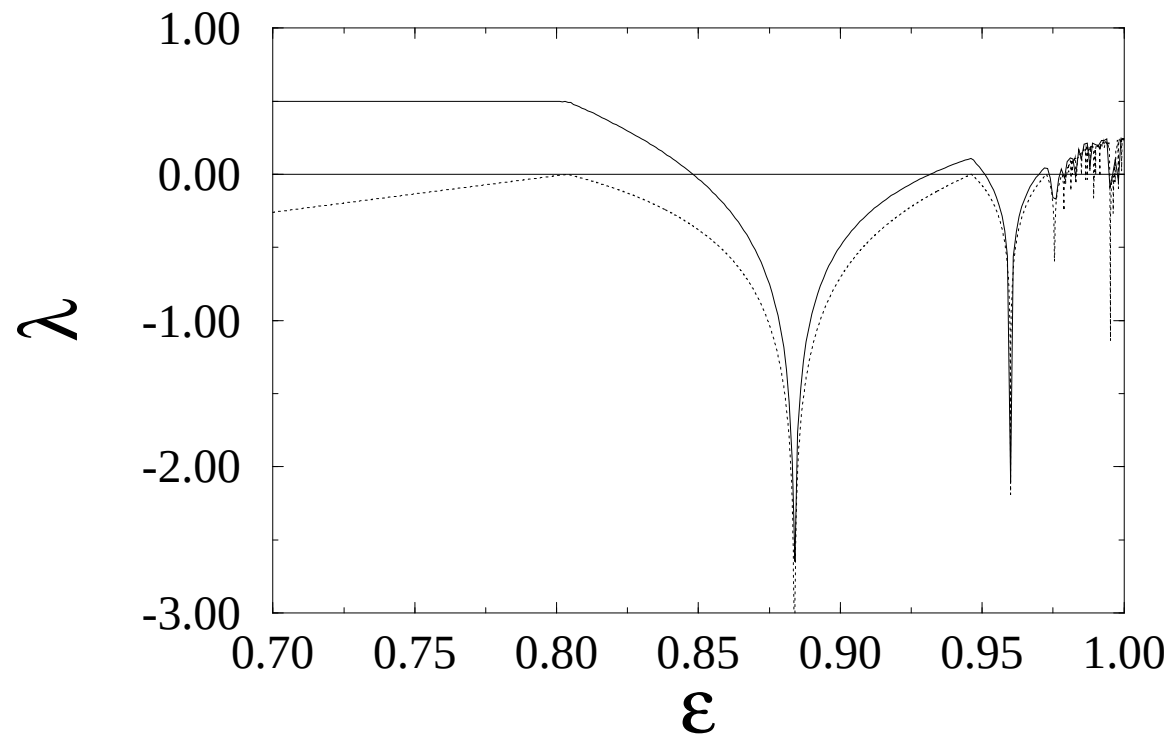

FIG. 5. Lyapunov exponent versus the coupling constant $\epsilon$. The nonlinearity is $\alpha = 1.5$. Solid line: co-moving frame. Dotted line: stationary frame.

Even though convective instability eventually seems to lead to temporal bifurcations, these may occur only very far downflow. For large $\epsilon$, it was, e.g., verified that the temporal periodicity never becomes very high in a lattice of ten thousand sites. In a practical sense, therefore, spatial chaos with temporal periodicity may very well occur in a limited region of the lattice. The main difference with the convectively stable case lies in the stability against noise. If the pattern is convectively unstable, even the tiniest amount of noise will be amplified, and the pattern destroyed. This can also be understood from a more fundamental point of view. Chaos is sensitively dependent on initial conditions, hence any change in a lattice site which is part of a spatially chaotic region must lead to a completely different spatial pattern. When noise is present, the pattern needs to be different at every time step and temporal periodicity is impossible. Although one might expect some differences in the spatio-temporal chaos resulting from a noise source and the intrinsic spatio-temporal chaos in the STP region, such differences could not be found so far.

In a periodicity $k$ spatial map, several attractors may coexist. It is, e.g., very well possible that (at least in some parameter regions) the $k = 1$ attractor is also an attractor of the $k = 2$ spatial map. On the basis of preliminary data we would like to conjecture that the OCLL selects the solution which minimizes the co-moving Lyapunov exponent.

In conclusion, we found spatial chaos with temporal periodicity to exist in the one-way coupled logistic lattice. With the help of a newly introduced spatial map which can exactly reproduce the spatial patterns of the spatio-temporal coupled map lattice, it was not only shown that spatial chaos is convectively unstable, but also that stable patterns can exist arbitrarily close to chaos. By employing the concept of co-moving Lyapunov exponents, we furthermore found that an important aspect of the bifurcation behavior of the OCLL can be predicted on the basis of the corresponding spatial map.

**Acknowledgements**

We would like to thank K. Tokita and T. Yamamoto for fruitful discussions. This work was supported by a Grant-in-Aid of the Japan Society for the Promotion of Science (JSPS) under grant no. 93043, and a Grant-in-Aid for Scientific Research from the Ministry of Educations, Science and Culture, Japan under grant no. 05836006.

* e-mail: frederik@complex.c.u-tokyo.ac.jp

† e-mail: kaneko@cyber.c.u-tokyo.ac.jp